\documentclass[aps,twocolumn,showpacs,preprintnumbers,amsmath,amssymb,superscriptaddress,floatfix,nofootinbib]{revtex4}

\usepackage{graphicx}
\usepackage{epsfig}
\usepackage{epstopdf}
\usepackage{hyperref}
\usepackage{amsmath}
\usepackage{amsfonts}
\usepackage{amssymb}
\usepackage{color}

\begin{document}

\title{Study of the $f_2(1270)$, $f'_2(1525)$, $\bar{K}^*_{2}(1430)$, $f_0(1370)$ and $f_0(1710)$ production from $\psi (nS)$ and $\Upsilon (nS)$ decays}

\date{\today}

\author{Lian-Rong Dai}
\affiliation{Department of Physics, Liaoning Normal University,
Dalian 116029, China}\affiliation{Institute of Modern Physics, Chinese Academy of
Sciences, Lanzhou 730000, China}

\author{Ju-Jun Xie} \email{xiejujun@impcas.ac.cn}
\affiliation{Institute of Modern Physics, Chinese Academy of
Sciences, Lanzhou 730000, China} \affiliation{Research Center for
Hadron and CSR Physics, Institute of Modern Physics of CAS and
Lanzhou University, Lanzhou 730000, China} \affiliation{State Key
Laboratory of Theoretical Physics, Institute of Theoretical Physics,
Chinese Academy of Sciences, Beijing 100190, China}

\author{Eulogio Oset}
\affiliation{Institute of Modern Physics, Chinese Academy of
Sciences, Lanzhou 730000, China} \affiliation{Departamento de
F\'{\i}sica Te\'orica and IFIC, Centro Mixto Universidad de
Valencia-CSIC Institutos de Investigaci\'on de Paterna, Apartado
22085, 46071 Valencia, Spain}

\begin{abstract}

Based on previous studies that support the important role of the
$f_2(1270)$, $f'_2(1525)$, and $\bar{K}^{*}_{2}(1430)$ resonances in
the $J/\psi [\psi(2S)] \to \phi (\omega) VV$ decays, we make an
analysis of the analogous decays of $\Upsilon (1S)$ and $\Upsilon
(2S)$, taking into account recent experimental data. In addition, we
study the $J/\psi$ and $\psi(2S)$ radiative decays and we also made
predictions for the radiative decay of $\Upsilon(1S)$ and
$\Upsilon(2S)$ into $\gamma f_2(1270)$, $\gamma f'_2(1525)$, $\gamma
f_0(1370)$ and $\gamma f_0(1710)$, comparing with the recent results
of a CLEO experiment. We can compare our results for ratios of decay
rates with eight experimental ratios and find agreement in all but
one case, where experimental problems are discussed.
\end{abstract}
 \maketitle

\section{Introduction}

Properties of mesons are key issues for understanding the
confinement mechanism of QCD. Within the traditional constituent
quark models, mesons are described as quark-antiquark ($q\bar{q}$)
states. This picture could explain successfully the properties of
the ground states of the flavor SU(3) vector meson nonet. However,
there are many meson (or mesonlike) states that could not be
explained as $q\bar{q}$ states. Depending on their coupling to
specific production mechanisms and their decay pattern, these states
are interpreted as molecular-type excitations or as tetraquark
states. But, there is debate on their exotic structure, unlike for
states that carry spin-exotic quantum numbers, e.g. $J^{PC}=1^{-+}$,
and hence cannot be $q\bar{q}$ states. One has an example in the
sector of light mesons with mass below $1$ GeV, where the scalar
mesons $\sigma(500)$, $a_0(980)$ and
$f_0(980)$~\cite{Agashe:2014kda} have been largely debated. Long ago
it was suggested that the $f_0(980)$ and $a_0(980)$ resonances could
be weakly bound states of $K \bar K$ \cite{Weinstein:1982gc}. The
advent of chiral unitary theory has brought new light into this
issue and by now the  $\sigma(500)$, $a_0(980)$ and $f_0(980)$ are
accepted as dynamically generated states from the interaction of
coupled channels $\pi \pi, K \bar K, \eta \eta, \pi \eta$ in $S$
wave~\cite{Oller:2000ma,npa,ramonet,kaiser,markushin,juanito,rios}.

Similarly, in Refs.~\cite{Geng:2008gx,Geng:2009gb}, the former work
of Ref.~\cite{Molina:2008jw} on the $\rho \rho$ interaction was
extended to SU(3) using the local hidden gauge formalism for
vector-vector interaction and a unitary approach in coupled
channels. This interaction generates resonances, some of which can
be associated to known resonances, namely the $f_2(1270)$,
$f'_2(1525)$, and $\bar{K}^{*}_{2}(1430)$, as well as the
$f_0(1370)$ and $f_0(1710)$. The results obtained in those former
works gave support to the idea of the $f_2(1270)$, $f'_2(1525)$,
$\bar{K}^{*}_{2}(1430)$, $f_0(1370)$ and $f_0(1710)$ as being
quasimolecular states of two vector mesons. In reactions producing
these resonances, a pair of vector mesons are primary produced and
these two vector mesons rescatter after the production, giving rise
to the resonances that can be observed in the invariant mass
distributions.

In Ref.~\cite{daizou}, the important role of the $f_2(1270)$,
$f'_2(1525)$, and $\bar{K}^{*}_{2}(1430)$ in the $J/\psi \to \phi
(\omega) VV$ decays was studied based on the vector-vector molecular
structure of those resonances. Related work was also done in
Ref.~\cite{hanhart} interpreting the $J/\psi$ radiative decay into
these resonances. Those latter works were then extended in
Ref.~\cite{Dai:2013uua} to study the decay of $\psi(2S)$ into
$\omega (\phi)$ and $f_2(1270)$, $f'_2(1525)$ and $\psi(2S)$ into
$K^{*\,0}$ and $\bar{K}^{*\,0}_{2}(1430)$. At the same time, in this
latter work the ideas of Ref.~\cite{hanhart} in the radiative decay
were extended to the decay of $\Upsilon (1S)$, $\Upsilon (2S)$ and
$\psi(2S)$. These hadronic and  radiative decays for $J/\psi$ have
also been addressed within a scheme where the $f_0(1370)$,
$f_0(1500)$ and $f_0(1710)$ states emerge as a result of glueball
quarkonia mixing~\cite{gutsche}. With the steady accumulation of
experimental data, new results are now available that can test these
theoretical ideas  and an update of the theoretical predictions has
become timely. In particular the very recent CLEO data on $J/\psi$,
$\psi (2S)$ and $\Upsilon(1S)$ radiative decays~\cite{seth} are most
welcome.

In the present work, we make a reanalysis of those decays taking
into account the recent report of the CLEO data. In addition to the
$J/\psi$ and $\psi(2S)$ decays, we also made predictions for the
radiative decay of $\Upsilon(1S)$ and $\Upsilon(2S)$ into $\phi
(\omega) f_2(1270)$, $\phi (\omega) f'_2(1525)$, or $K^{*\,0}$ and
$\bar{K}^{*\,0}_{2}(1430)$. We evaluate ratios of decay rates and
can compare with eight experimental ratios. The agreement found with
experiment is good, with one exception that will require future test
due to present experimental difficulties.

\section{Hadronic decay}

\subsection{Formalism for $\Upsilon (1S)$ decay into  $\omega(\phi) VV$}

\begin{figure*}[htbp]
\begin{center}
\includegraphics[scale=0.8]{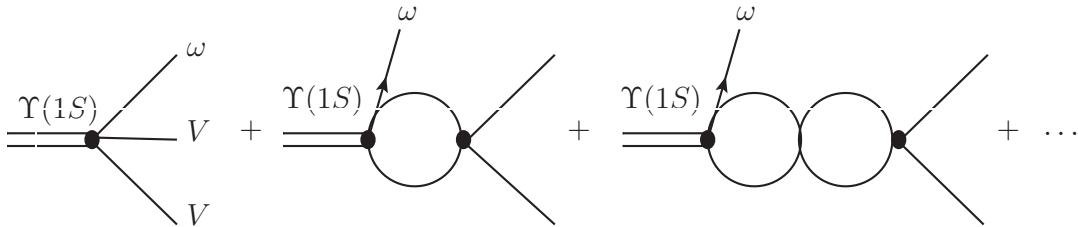}
\caption{Production mechanism of $\omega$ and two interacting vector
mesons.} \label{Fig1}
\end{center}
\end{figure*}

We extend here the formalism used in Refs.~\cite{daizou,Dai:2013uua}
to study the decay of $\Upsilon (1S)$  into $\omega(\phi)$ and two
interacting vectors, $V V$, that lead to the tensor state. The
mechanism is depicted in Fig.~\ref{Fig1}. We follow the approach of
Ref.~\cite{Roca:2004uc} and write the $\phi$ and $\omega$ as a
combination of a singlet and an octet of SU(3) states
\begin{eqnarray}
\omega&=&\frac{1}{\sqrt{2}}(u\bar{u}+d\bar{d})=\sqrt{\frac{2}{3}}V_1+\sqrt{\frac{1}{3}}V_8 \nonumber \\
\phi&=&s\bar{s}=\sqrt{\frac{1}{3}}V_1-\sqrt{\frac{2}{3}}V_8
\end{eqnarray}

The two $VV$ states combine to $I=0$, either with a
$\frac{1}{\sqrt{2}}(u\bar{u}+d\bar{d})$ or $s\bar{s}$ SU(3)
structure to match the SU(3) singlet nature of the $b \bar b$ state.
One obtains matrix elements for the  $\Upsilon(1S) \to \omega
\frac{1}{\sqrt{2}}(u\bar{u}+d\bar{d}) \to \omega VV$ and
$\Upsilon(1S) \to \omega s\bar{s} \to \omega VV$ amplitudes with the
results
\begin{eqnarray}
\frac{2}{3}T^{(1,1)}+\frac{1}{3}T^{(8,8)} \quad\quad
\mbox{and}\quad\quad
\frac{\sqrt{2}}{3}T^{(1,1)}-\frac{\sqrt{2}}{3}T^{(8,8)}
\end{eqnarray}
respectively, where $T^{(1,1)}$ is the $T$ matrix for the singlet of
$\phi$ and the one of $VV$ giving the vacuum and $T^{(8,8)}$ the
corresponding part for the octet.

Similarly, for the $\Upsilon(1S) \to \phi
\frac{1}{\sqrt{2}}(u\bar{u}+d\bar{d}) \to \phi VV$ and $\Upsilon(1S)
\to \phi s\bar{s} \to \phi VV$, we obtain
\begin{eqnarray}
\frac{\sqrt{2}}{3}T^{(1,1)}-\frac{\sqrt{2}}{3}T^{(8,8)}
\quad\quad\mbox{and}\quad\quad
\frac{1}{3}T^{(1,1)}+\frac{2}{3}T^{(8,8)}
\end{eqnarray}
respectively.

It was found in Ref.~\cite{daizou} that in terms of $VV$ the
$\frac{1}{\sqrt{2}}(u\bar{u}+d\bar{d})$ and $s\bar{s}$ components
could be written as
\begin{eqnarray}
\frac{1}{\sqrt{2}}(u\bar{u}+d\bar{d}) \to
\frac{1}{\sqrt{2}}(\rho^0\rho^0+\rho^+\rho^-+\rho^-\rho^+ + \omega \omega  \nonumber \\
+ K^{*+}K^{*-} + K^{*0}\bar{K}^{*0}), \\
s\bar{s} \to K^{*-}K^{*+}+\bar{K}^{*0}K^{*0}+\phi\phi.
\end{eqnarray}

\begin{figure}[htbp]
\begin{center}
\includegraphics[scale=1.]{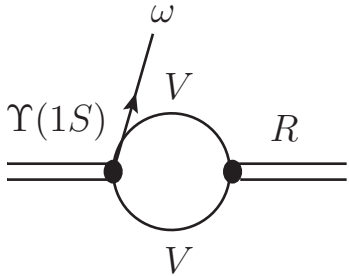}
\caption{Selection of diagrams of Fig.~\ref{Fig1} that go into
resonance formation, omitting the coupling to $V V$ without
interaction.} \label{Fig2}
\end{center}
\end{figure}

Only the terms in Fig.~\ref{Fig1} where the $VV$ interact lead to
the tensor resonance. Then we remove the first diagram of
Fig.~\ref{Fig1} corresponding to the tree level and this leads to
the diagram depicted in Fig.~\ref{Fig2}.

The final transition matrix for $\Upsilon (1S) \to \omega(\phi,{K^*}^0)R$,
with $R$ the resonance under consideration, is given by
\begin{eqnarray}
t_{\Upsilon (1S)\to \omega R}=\sum_{j}W_{j}^{(\omega)}G_{j}g_{j} ,
\label{tJ}
\end{eqnarray}
with $W_{j}^{(\omega)}$ the weights given in Ref.~\cite{daizou},
$G_{j}$ the $VV$ loop functions and $g_{j}$ the couplings of the
resonance considered to the corresponding $VV$ channel $j$. We
proceed similarly for $\Upsilon(1S) \to \phi R$ or $\Upsilon(1S) \to
{K^*}^0 R$ and all values of $W, G, g$ are tabulated in
Ref.~\cite{daizou}.

The $\Upsilon(1S)$ partial decay width is then given by
\begin{eqnarray}
\Gamma=\frac{1}{8\pi}\frac{1}{M^2_{\Upsilon (1S)}}|t|^2q \label{Gama}
\end{eqnarray}
with $q$ the momentum of the $\omega (\phi, K^{*\,0})$ in the
$\Upsilon(1S)$ rest frame.

We apply the formalism to the same decay channels but from the
$\Upsilon(2S)$ state. In terms of quarks the $\Upsilon(2S)$ state is
the $b\bar{b}$ state with the same quantum numbers as the $\Upsilon
(1S)$, with a radial excitation to the $2S$ state of a $b\bar{b}$
potential. We then expect the same behavior as for the
$\Upsilon(1S)$, which stands as the $1S$ $b\bar{b}$ state. In the
derivation we have only used two properties from the $\Upsilon (1S)$
concerning these decays. First that it is an SU(3) singlet, which
can also be said of the  $\Upsilon (2S)$ state. The other dynamical
feature is related to the Okubo-Zweig-Iizuka (OZI) violation and the
weight going into $\phi s\bar{s}$ or
$\phi\frac{1}{\sqrt{2}}(u\bar{u}+d\bar{d})$ in its decay, that we
parametrize in terms of  $\nu =
T^{(1,1)}/T^{(8,8)}$~\cite{daizou,Dai:2013uua}. Given the fact that
this is also a dynamical feature not related to the internal
excitation of the $b\bar{b}$ quarks in the potential well, we shall
also assume that the  $\nu$ parameter is the same for  $\Upsilon
(2S)$  as for $\Upsilon (1S)$. The normalization of the
$W_{j}^{(\omega)}$ weights can be different but this will cancel in
the ratios. With these two reasonable assumptions we can make
predictions for the following four ratios that are discussed in the
next subsection.

\subsection{Numerical results of hadronic decay}

We collect the new results on these two sets of reactions, which are
$\Upsilon(1S)$ and $\Upsilon(2S)$ hadronic decays, respectively.

In Ref.~\cite{Dai:2013uua} results were given for the ratios
$\widehat{R}_1$, $\widehat{R}_2$, $\widehat{R}_3$, and
$\widehat{R}_4$, for $\psi(2S) \to \omega (\phi, K^{*\,0}) R$, and
the ratios were found compatible with experiments. We generalize
them here to the $\Upsilon(1S)$ decay.
\begin{eqnarray}
\widehat{R_{1}} & \equiv & \frac{\Gamma_{\Upsilon (1S)\to\phi
f_{2}(1270)}}{\Gamma_{\Upsilon (1S)\to\phi f^\prime_{2}(1525)}}, \\
\widehat{R_{2}} & \equiv & \frac{\Gamma_{\Upsilon (1S)\to\omega
f_{2}(1270)}}{\Gamma_{\Upsilon (1S)\to\omega f^\prime_{2}(1525)}},
\\
\widehat{R_{3}} & \equiv & \frac{\Gamma_{\Upsilon (1S)\to\omega
f_{2}(1270)}}{\Gamma_{\Upsilon (1S)\to\phi f_{2}(1270)}}, \\
\widehat{R_{4}} & \equiv & \frac{\Gamma_{\Upsilon (1S)\to K^{*0}
\bar{K}^{*0}_{2}(1430)}}{\Gamma_{\Upsilon (1S)\to\omega
f_{2}(1270)}}. \label{Upsilon1S}
\end{eqnarray}
Here we bring new data for new decays reported in
Ref.~\cite{Shen:2012iq} on $\Upsilon(1S)$ and $\Upsilon(2S)$ decays.
Concretely, $Br[\Upsilon(1S) \to \phi f'_2(1525)]$, $Br[\Upsilon(1S)
\to \omega f_2(1270)]$, $Br[\Upsilon(1S) \to K^{*0}(892)
\bar{K}^{*0}_2(1430)]$, and the same decays for $\Upsilon(2S)$. In
Table~\ref{Tab:upsilon1sdecays} we show the new numbers for the
$\Upsilon(1S)$ decays. The criteria used to obtain the theoretical
errors are the same as in Ref.~\cite{Dai:2013uua}. For this case, we
have three data and can obtain two ratios, and as we can see, we
find agreement with experiment within errors.

\begin{table}[htbp]
\caption{Numerical results of $\Upsilon (1S)$ decays.
$\widehat{R_{1}}\cdot\widehat{R_{3}}$ provides the ratio
$\Gamma_{\Upsilon (1S)\to\omega f_{2}(1270)}/\Gamma_{\Upsilon
(1S)\to\phi f^\prime_{2}(1525)}$. Experimental data are taken from
Ref.~\cite{Shen:2012iq}. The numbers in parentheses are the
theoretical values with upper and lower errors. The numbers before
them indicate the band of theoretical values considering the former
errors.}
\begin{center}
\begin{tabular}{ccc}
\hline\hline
& Theory~~~ & ~~~~~~Experiment ~~~~~~ \\
\hline
\\
$\widehat{R_{1}}$ & $0.11-0.52 ~(0.24^{+0.28}_{-0.13})$  &\\[1.5ex]
$\widehat{R_{2}}$ & $2.58-11.99 ~(5.19^{+6.80}_{-2.61})$ &\\[1.5ex]
$\widehat{R_{3}}$ & $5.64-17.46 ~(9.67^{+7.77}_{-4.06})$ &\\[1.5ex]
$\widehat{R_{4}}$ & $0.94-2.37 ~(1.50^{+0.87}_{-0.56})$ & 1.75-10.9 \\[1.5ex]
$\widehat{R_{1}}\cdot\widehat{R_{3}}~~~$ &  $0.73-5.61 ~(2.32^{+3.29}_{-1.59})$ & 0.0-8.76\\[1.5ex]
\hline \hline
\end{tabular}
\end{center} \label{Tab:upsilon1sdecays}
\end{table}

For the case of $\Upsilon(2S)$ hadronic decays, we define the
equivalent four ratios,
\begin{eqnarray}
\overline{R_{1}} & \equiv & \frac{\Gamma_{\Upsilon (2S)\to\phi
f_{2}(1270)}}{\Gamma_{\Upsilon (2S)\to\phi f^\prime_{2}(1525)}},\\
\overline{R_{2}} & \equiv & \frac{\Gamma_{\Upsilon (2S)\to\omega
f_{2}(1270)}}{\Gamma_{\Upsilon (2S)\to\omega f^\prime_{2}(1525)}},
\\
\overline{R_{3}} & \equiv & \frac{\Gamma_{\Upsilon (2S)\to\omega
f_{2}(1270)}}{\Gamma_{\Upsilon (2S)\to\phi f_{2}(1270)}}, \\
\overline{R_{4}} & \equiv & \frac{\Gamma_{\Upsilon (2S)\to K^{*\,0}
\bar{K}^{*\,0}_{2}(1430)}}{\Gamma_{\Upsilon (2S)\to\omega
f_{2}(1270)}}. \label{Upsilon2S}
\end{eqnarray}

The values of these ratios are shown in
Table~\ref{Tab:upsilon2sdecays}. For the case of $\Upsilon(2S)$ the
datum for $\omega f_2(1270)$ with negative width and large errors
cannot be used for ratios and hence we can only construct one ratio.
We can see that we find agreement of the theoretical numbers with
the only experimental ratio that we can form.

\begin{table}[htbp]
\caption{Numerical results of $\Upsilon (2S)$ decays.
$\overline{R_{1}}\cdot\overline{R_{3}}\cdot\overline{R_{4}}$
provides the ratio $\Gamma_{\Upsilon (2S)\to K^{*\,0}
\bar{K}^{*\,0}_{2}(1430)}/\Gamma_{\Upsilon (2S)\to\phi
f^\prime_{2}(1525)}$. Experimental data are taken from
Ref.~\cite{Shen:2012iq}. The numbers in parentheses are the
theoretical values with upper and lower errors. The numbers before
them indicate the band of theoretical values considering the former
errors.}
\begin{center}
\begin{tabular}{ccc}
\hline\hline
& Theory~~~ &~~~~~~ Experiment~~~~~~ \\
\hline
\\
$\overline{R_{1}}$ & $0.11 -  0.51 ~(0.24^{+0.27}_{-0.13})$ & \\[1.5ex]
$\overline{R_{2}}$ & $2.58 - 11.99 ~(5.19^{+6.79}_{-2.61})$ & \\[1.5ex]
$\overline{R_{3}}$ & $5.63 - 17.45 ~(9.69^{+7.76}_{-4.06})$ & \\[1.5ex]
$\overline{R_{4}}$ & $0.94 -  2.37 ~(1.50^{+0.87}_{-0.56})$ & \\[1.5ex]
$\overline{R_{1}}\cdot\overline{R_{3}}\cdot\overline{R_{4}}~~~$ & $0.77 - 8.71 ~(3.49^{+5.22}_{-2.72})$ & $0.0 - 7.92$  \\[1.5ex]
\hline \hline
\end{tabular}
\end{center} \label{Tab:upsilon2sdecays}
\end{table}

\section{Radiative decay}

\subsection{Formalism for $\psi(nS)$ and $\Upsilon (nS)$ decay into $\gamma VV$ }

\begin{figure*}[htbp]
\begin{center}
\includegraphics[scale=0.6]{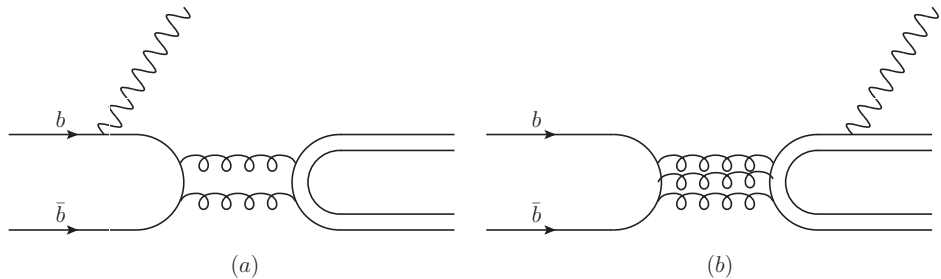}
\caption{Two mechanisms of the $\Upsilon(1S)$ radiative decays.}
\label{Fig3}
\end{center}
\end{figure*}

Another successful test on the vector-vector nature of the
$f_2(1270)$ and $f'_2(1525)$ was done in Ref.~\cite{hanhart} by
looking at the  decay of $J/\psi$  into $\gamma T$, where T is any
of these two tensor resonances. A justification was given in
Ref.~\cite{hanhart} for the photon being radiated from the initial
$c\bar{c}$ state. The remaining $c\bar{c}$ gave rise to a pair of
vector mesons which upon rescattering produced the tensor
resonances. The only dynamical assumption made in
Ref.~\cite{hanhart} was that the photon was radiated from the
$c\bar{c}$ state, not from the final $VV$ state. Translated to the
present problem, the argument is based on the dominance of the
diagram of Fig.~\ref{Fig3}(a) over the one of Fig.~\ref{Fig3}(b),
which require two, versus three, gluon exchange as discussed in
Ref.~\cite{Kopke:1988cs} for the $J/\psi$ case.

\begin{figure*}[htbp]
\begin{center}
\includegraphics[scale=0.9]{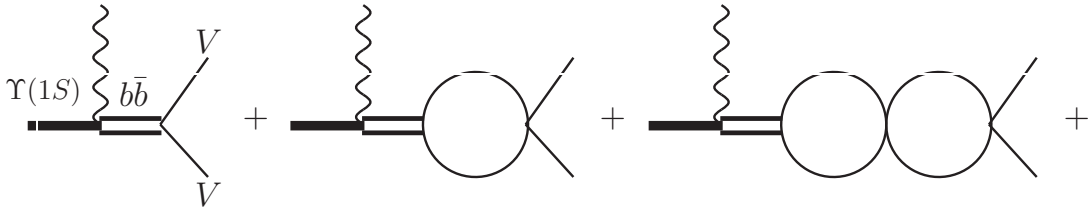}
\caption{Schematic representation of $\Upsilon (1S)$ decay into a
photon and one dynamically generated resonance.} \label{Fig4}
\end{center}
\end{figure*}

In the present work the $b\bar{b}$ state is assumed to be an SU(3)
singlet in $\Upsilon$ decay, like in the case of the $c\bar{c}$
state. Both assumptions hold equally here and hence the only
difference in the results stems from an overall normalization, which
disappears when ratios are made, and the momenta $q$ in the formula
of the width, since now the $\Upsilon$ mass is different to the one
of the $J/\psi$. It is easy to extend the results of $J/\psi$ to the
decay of the $\Upsilon$. The mechanism is depicted in
Fig.~\ref{Fig4} for $\Upsilon (1S)$ radiative decay. When this is
taken into account, we can evaluate the same ratios for
$\Upsilon(nS)$ radiative decay as those for $J/\psi$ case.

The extended formalism for the transition amplitudes of
$\Upsilon(1S)$ decay into $\gamma T$ provides the amplitudes
\begin{eqnarray}\label{radi}
t_{\Upsilon (1S)\rightarrow \gamma \mathrm{R}}=\sum_j
\widetilde{w_j} G_j g_j \ .
\end{eqnarray}
and the weights $\widetilde{w_j}$ are the same as those obtained in
Ref.~\cite{hanhart} and given by
\begin{eqnarray}\label{eq:cg}
\widetilde{ w_i} = c~\left\{\begin{array}{ll}
             -\sqrt{\frac{3}{2}}&\quad\mbox{for $\rho\rho$}\\
             -\sqrt{2}&\quad\mbox{for $K^*\bar{K}^*$}\\
             \frac{1}{\sqrt{2}} &\quad \mbox{for $\omega\omega$}\\
             \frac{1}{\sqrt{2}} &\quad\mbox{for $\phi\phi$}
            \end{array}
\right. .
\end{eqnarray}
where $c$ is a normalization constant, which cancels in the ratios,
and $G_j$, $g_j$ are again the loop functions of the intermediate
$VV$ states and the couplings of the resonance to these $VV$
channels. All these quantities are given in Table 1 of
Ref.~\cite{hanhart}. The same theoretical framework allows us to
evaluate the $\Upsilon (1S)$ radiative decay into the scalar meson
$f_0(1370)$ and $f_0(1710)$ which are also obtained from the
interaction of $VV$, mostly $\rho\rho$ and $K^{*0} \bar{K}^{*\,0}$
respectively. The decay width is given again by Eq. (\ref{Gama})
where $q$ is the momentum of the photon in the $\Upsilon (1S)$  rest
frame.

\subsection{Numerical results of radiative decays}

For the radiative decays, we find in the PDG new results for
$\Upsilon(1S)$. For the $\Upsilon(2S)$ there are only upper bounds
and we cannot compare ratios. There is also a new set of data on
$J/\psi \to \gamma T$ and $\psi(2S) \to \gamma T$ from
Ref.~\cite{seth}, and similarly going to $\gamma S$, where $S$ is any of the
scalar mesons $f_0(1370)$, $f_0(1710)$.

We have taken advantage of the fact that these data for $J/\psi$ and
$\psi(2S)$ radiative decays come from the same experiment, so they
are advantageous for the evaluation of ratios since they usually
cancel systematic uncertainties. The data of Ref.~\cite{seth} are
given for $J/\psi [\psi(2S)] \to \gamma R \to \gamma \pi \pi$ or
$J/\psi [\psi(2S)] \to \gamma R \to \gamma K \bar{K}$. In order to
convert those numbers in partial decay widths we divide by the $V
\to \pi \pi$ or $V \to K \bar{K}$ branching ratio and add relative
errors in quadrature. In Table~\ref{Tab:pipikkbr} we show the
branching ratios that we used in the present work. Some of these
branching ratios are not well known, and there are different values
for them, such as for $f_0(1370) \to K \bar{K}$ and $f_0(1710) \to K
\bar{K}$. We take an approximate average value, compatible with the
different results.

\begin{table*}[htbp]
\caption{Values of some of the parameters used in this work.}
\begin{tabular}{ccccc}
\hline \hline
& \multicolumn{4}{c}{Decay channels} \\
\hline    & $f_2(1270) \to \pi \pi$ ~~~ & $f'_2(1525) \to K \bar{K}$ ~~~ & $f_0(1370) \to K \bar{K}$ ~~~ & $f_0(1710) \to K \bar{K}$ \\
\hline  Ref. & \cite{Agashe:2014kda}   & \cite{Agashe:2014kda}  & \cite{Bugg:1996ki,Bugg:2007ja} ~~~~~ \cite{Albaladejo:2008qa} & \cite{Albaladejo:2008qa} ~~~~~~~ \cite{Longacre:1986fh} ~~~ \cite{Geng:2008gx} \\
        Branching ratios (\%) & $84.8^{+2.4}_{-1.2}$  & $88.7 \pm 2.2$ & $35 \pm 13$ ~~ $\sim 20$     &  $36 \pm 12$ ~~ $38^{+9}_{-19}$ ~~ $55$\\
        Adopted value (\%)    & $84.8 \pm 1.3$        & $88.7 \pm 2.2$ & $30 \pm 15$ & $35 \pm 15$ \\
\hline \hline
\end{tabular} \label{Tab:pipikkbr}
\end{table*}

Then we evaluate the ratios,
\begin{eqnarray}
R_T &=& \frac{\Gamma_\mathrm{J/\psi\rightarrow \gamma
f_2(1270)}}{\Gamma_{J/\psi\rightarrow \gamma f'_2(1525)}}, \\
R_S &=& \frac{\Gamma_{J/\psi\rightarrow \gamma
f_0(1370)}}{\Gamma_{J/\psi\rightarrow\gamma
f_0(1710)}}, \\
\widetilde{R_T} &=& \frac{\Gamma_\mathrm{\psi(2S)\rightarrow \gamma
f_2(1270)}}{\Gamma_{\psi (2S)\rightarrow \gamma f'_2(1525)}} , \\
\widetilde{R_S} &=& \frac{\Gamma_{\psi (2S)\rightarrow \gamma
f_0(1370)}}{\Gamma_{\psi(2S)\rightarrow\gamma
f_0(1710)}}, \\
\widehat{R_T} &=& \frac{\Gamma_\mathrm{\Upsilon(1S)\rightarrow
\gamma f_2(1270)}}{\Gamma_{\Upsilon(1S)\rightarrow \gamma
f'_2(1525)}} , \\
\widehat{R_S} &=& \frac{\Gamma_{\Upsilon(1S)\rightarrow \gamma
f_0(1370)}}{\Gamma_{\Upsilon(1S)\rightarrow\gamma
f_0(1710)}} ,\\
\overline{R_T} &=& \frac{\Gamma_\mathrm{\Upsilon(2S)\rightarrow
\gamma f_2(1270)}}{\Gamma_{\Upsilon(2S)\rightarrow \gamma
f'_2(1525)}} , \\
\overline{R_S} &=& \frac{\Gamma_{\Upsilon(2S)\rightarrow \gamma
f_0(1370)}}{\Gamma_{\Upsilon(2S)\rightarrow\gamma f_0(1710)}}.
\end{eqnarray}

The numerical results are summarized in Table~\ref{Tab:ratios}
compared with the experimental data. We note that the comparison
with the experimental results is particularly valuable since the
theoretical results were predictions done before (see
Ref.~\cite{Dai:2013uua}) that can be contrasted with data observed
later. We see that we have agreement in all numbers except for the
ratio of $\widetilde{R_T}$ where even within errors there is a
discrepancy of about a factor two. Actually, the reason for the
large experimental value of this ratio is the small value for
$Br[\psi(2S) \to \gamma f'_2(1525)]$. One can see in
Ref.~\cite{seth} that this rate is small in absolute value since the
ratio $\mathfrak{B}_2[\psi(2S)]/\mathfrak{B}_2[J/\psi]$ is of
$4.1\%$ (see more details of Table VI in Ref.~\cite{seth}) and is
the smallest one of the eight ratios tabulated there, diverging
significantly from the $13\%$ rule for this ratio. We have consulted
the authors of Ref.~\cite{seth}, who admit problems in this datum
for the $\psi(2S)$ transitions, as we can see in Fig. 9 of
Ref.~\cite{seth}, where the relative strength of $f'_2(1525)$ and
$f_0(1710)$ are quite different in the decay modes $\psi(2S) \to
\gamma K^+ K^-$ and $\psi(2S) \to \gamma K_s K_s$, when they should
be the same.

\begin{table*}[htpb]
\caption{Ratios of the molecular model and in comparison with data
from Refs.~\cite{CLEO1,CLEO2,CLEO3,seth}. \label{Tab:ratios}}
\begin{center}
\begin{tabular}{c|c|c}
\hline\hline & Molecular picture &  Data\\\hline
$R_T~(J/\psi)$ & $2\pm 1 $ (Ref.~\cite{hanhart}) & $3.18^{+0.58}_{-0.64}$ (Refs.~\cite{CLEO1,CLEO2,CLEO3})  $2.57 \pm 0.5$ (Ref.~\cite{seth}) \\
$R_S~(J/\psi)$ & $1.2\pm 0.3$ & $0.51 \pm 0.41 $ (Ref.~\cite{seth})  \\
\hline
$\widetilde {R_T}~(\psi (2S))$ & $1.94 \pm 0.97$ & $8.6 \pm 3.0$ (Ref.~\cite{seth}) \\
$\widetilde{R_S}~(\psi (2S))$ & $1.14\pm 0.28$ & $ 0.81^{+0.84}_{-0.73}$ (Ref.~\cite{seth}) \\
\hline
$\widehat{R_T}~(\Upsilon (1S))$ & $1.84\pm 0.92$ &  $2.66 \pm 0.67$ (Ref.~\cite{seth}) \\
$\widehat{R_S}~(\Upsilon (1S))$ & $1.05\pm 0.26$  & \\
\hline
$\overline{R_T}~(\Upsilon (2S))$ & $ 1.83\pm 0.92$ &  \\
$\overline{R_S}~(\Upsilon (2S))$ & $1.05\pm 0.26$ & \\
\hline\hline
\end{tabular}
\end{center}
\end{table*}

The problem for this $\psi(2S)$ transition, together with the
discrepancy with the theoretical results, which have otherwise been
successful in the other cases, should serve as incentive to have a
further experimental look into this transition.

\section{Conclusion}

A further test on the molecular nature of the $f_2(1270)$,
$f'_2(1525)$ and $K^*_2(1430)$ has been made, using the decay of
$\Upsilon(nS)$ into $\phi (\omega)$ and any of the $f_2(1270)$,
$f'_2(1525)$ resonances, or $K^*(892)$ and $K^*_2(1430)$. We have
also studied the same decays from the $\psi(nS)$ state. The theory
only makes use of the fact that both $\psi(nS)$ and $\Upsilon(nS)$
are singlets of SU(3). A dynamical factor for the OZI violation into
the strange and nonstrange sectors, the $\nu$ parameter, is taken
from other experiments. The needed modifications due to kinematics
with respect to the analogous cases of $J/\psi$, $\psi(2S)$ decays
have been done and results for the decays of the  $\Upsilon(nS)$ are
found in agreement with experiment for the cases where experimental
information is available.

We also analyzed the radiative decay $J/\psi$, $\psi(2S)$,
$\Upsilon(1S)$ and $\Upsilon(2S)$ into a photon and a tensor
$f_2(1270)$, $f'_2(1525)$ or a photon and a scalar $f_0(1370)$,
$f_0(1710)$. New data on these decays has been reported recently
from the CLEO Collaboration which has allowed us to compare with
predictions for these decays made prior to the experiment. The
agreement found with experiment is good in all cases except in one
ratio involving the $Br[\psi(2S) \to \gamma f'_2(1525)]$ decay which
was found exceptionally small in the experiment and was admitted as
problematic there. Those problems and the discrepancy with the
theory, which otherwise is in agreement with the data, calls for a
further reanalysis of this datum.

The overall agreement found with the data on different experiments
provides extra support for the picture in which the tensor states
$f_2(1270)$, $f'_2(1525)$, $\bar{K}^{*}_{2}(1430)$, as well as the
scalar ones $f_0(1370)$ and $f_0(1710)$ are dynamically generated
states from the vector-vector interaction.

\section*{Acknowledgments}

We would like to thank K.K. Seth for calling our attention to the
experiment of Ref.~\cite{seth} and for very useful discussions. One
of us, E. O., wishes to acknowledge support from the Chinese Academy
of Science in the Program of Visiting Professorship for Senior
International Scientists (Grant No. 2013T2J0012). This work is
partly supported by the Spanish Ministerio de Economia y
Competitividad and European FEDER funds under the contract number
FIS2011-28853-C02-01 and FIS2011-28853-C02-02, and the Generalitat
Valenciana in the program Prometeo, 2009/090. We acknowledge the
support of the European Community-Research Infrastructure
Integrating Activity Study of Strongly Interacting Matter (acronym
HadronPhysics3, Grant Agreement n. 283286) under the Seventh
Framework Programme of EU. This work is also partly supported by the
National Natural Science Foundation of China under Grant Nos.
11475227, 11375080, and 10975068, and the Natural Science Foundation
of Liaoning Scientific Committee under Grant No. 2013020091.

\end{document}